     \documentclass[12pt]{article}
     \usepackage[margin=1in]{geometry}
     \usepackage{times}
     \usepackage{lipsum}
\usepackage{setspace}
\usepackage{epigraph}

\usepackage{sectsty}
\usepackage[italicdiff]{physics}

\usepackage{amsmath}

\usepackage [english]{babel}

\usepackage{etex}
\usepackage{amsthm}
\usepackage{amssymb}
\usepackage{soul}
\usepackage{calc}
\usepackage[usenames,dvipsnames]{xcolor}
\usepackage{mathptmx}
\usepackage{colortbl}
\usepackage[boxed]{algorithm2e}
\usepackage{epstopdf}
\usepackage{arydshln}
\usepackage{booktabs}
\usepackage{natbib}
\usepackage{hyperref}
\hypersetup{
	colorlinks=true,         
	linkcolor=MidnightBlue,          
	citecolor=MidnightBlue,
	urlcolor=MidnightBlue            
 }

\usepackage{breakurl}
\usepackage{bookmark}
\usepackage{etoolbox}
\usepackage{graphicx}
\theoremstyle{plain}
\theoremstyle{definition}

\usepackage[title]{appendix}
\usepackage{csquotes}
\usepackage{comment}

\usepackage{authblk}
\newcommand{\be}{\begin{equation}}
\newcommand{\ee}{\end{equation}}

\title{What Is a Reference Frame in General Relativity?}
\author[1,2]{Nicola Bamonti}
\date{ }

\affil[1]{Department of Philosophy, Scuola Normale Superiore, Piazza dei Cavalieri, 7, Pisa, 56126, Italy}
\affil[2]{Department of Philosophy, University of Geneva, 5 rue de Candolle, 1211 Geneva 4, Switzerland}

\begin{document}
\maketitle

%9812 WORDS

\begin{abstract}
\singlespacing
This work introduces a novel three‐fold classification of reference frames in General Relativity, distinguishing between Idealised Reference Frames (\textbf{IRFs}), Dynamical Reference Frames (\textbf{DRFs}), and Real Reference Frames (\textbf{RRFs}).
By defining a reference frame as a set of degrees of freedom instantiated by a physical system, the work contrasts this notion with that of \textit{coordinate systems}—purely mathematical idealisations lacking physical instantiation.
This classification addresses two longstanding challenges in GR: \textbf{(P1)} the difficulty of defining local and gauge–invariant observables, and \textbf{(P2)} how to interpret diffeomorphism gauge freedom in physical terms rather than as merely a mathematical redundancy. 
Overall, this work clarifies the conceptual foundations in classical GR, enhancing our understanding of gauge-symmetries, observers and laying the groundwork for future investigations in both classical and quantum gravitational contexts.
\end{abstract}

%\epigraph{Linguistic imprecision is the bane of good philosophy.}{K.P.Y. Thébault}

\clearpage

\tableofcontents

\clearpage

\section{Introduction} \label{Introduction}

In General Relativity (GR) the interplay between gauge symmetries and the physical interpretation of reference frames is of paramount importance.
Gauge transformations are transformations that lead to redundant descriptions of physical states. This redundancy, typically regarded as mathematical, complicates the identification of physically meaningful quantities.
In the Hamiltonian formulation these redundancies appear as first-class constraints, so that only quantities invariant under gauge transformations qualify as genuine \emph{observables} \citep{dirac:1950,Dirac:1958a,Dirac:1964}.
In this context an observable must commute with all constraints, ensuring that its Poisson bracket with any constraint (which generates an infinitesimal gauge transformation) vanishes. Thus, commutation ensures gauge invariance. 
In this work, I will consider the gauge group of GR the four-dimensional group of \textit{active} diffeomorphisms, $\text{Diff}(\mathcal{M})$, which \enquote*{move points around} \cite[p.170]{Isham:1992ms}, although this interpretation remains contested.\footnote{For an introduction to the distinction between active and passive diffeomorphisms, see \cite{RovelliGaul}. For a critique of this distinction, refer to \cite[p.14]{Weatherall2018-WEARTH-2}. The idea that active diffeomorphisms straightforwardly qualify as gauge symmetries in GR is not universally accepted (see, e.g. \citealp{Belot2017}).}
Although this definition is formally precise, identifying explicit observables in GR remains a formidable challenge, particularly when we require such observables to be local. 

A significant step toward resolving this challenge was taken by \cite{Rovelli_2002}, who distinguished between \emph{partial observables}—measurable quantities that need not be gauge-invariant—and \emph{complete observables} that result from combining partial observables in a gauge-invariant fashion.
Complete observables, that capture the theory's physical predictions, correspond to Dirac observables due to the correspondence between the Hamiltonian and covariant formulations \citep{Lee1990, Dittrich2006,Dittrich2007}, while, partial observables retain physical relevance as they characterise the actual measurements performed in experiments serving as the \enquote*{handles through which systems can couple} \citep{Rovelli2014}.\footnote{For further discussion on \enquote*{observables} in GR, see \cite{Bergmann1961}, \cite{Gryb2016}, and \cite{Pitts2022}.}
This relational perspective reveals that the true physical content of GR resides not in assignments to abstract manifold points, but in the inter-relations among dynamically coupled material degrees of freedom.

This observation is particularly striking when one considers that geometrical objects such as the metric $g_{ab}(p)$ depend on the manifold points $p \in \mathcal{M}$ and are therefore not gauge-invariant—a fact that epitomises the \emph{problem of  local observables} in GR.
As this paper will demonstrate, a natural resolution of this problem is achieved by localising physical quantities with respect to spatiotemporal \emph{reference frames} rather than through coordinates.

Historically, the terms \enquote*{reference frame} and \enquote*{coordinate system} have often been used interchangeably, a practice traceable to Einstein’s own work. This usage has been critically examined, particularly in its historical and philosophical dimensions by \cite{Norton1989, Norton1993-NORGCA}.\footnote{\label{1}According to \citeauthor{Norton1989}'s analysis, \lq{}Einstein's coordinates\rq{} possessed physical significance as structures in $\mathbb{R}^4$, termed the \enquote*{Einstein's manifold}. Modern practice, however, has interpreted these coordinates as \enquote*{coordinate charts}, referred to here as coordinate systems, which are merely labels for geometric structures on a smooth manifold $\mathcal{M}$, lacking any physical \textit{instantiation}. In this work, \enquote*{instantiation} refers to the representational relationship between a model and the physical possibility it aims to describe. For further discussion, see \cite{GomesSymmetries}.}
However, it is useful—especially for questions of observability—to keep explicit the conceptual distinction between coordinate systems and reference frames.

This paper clarifies the concept of reference frames in GR and their role in defining local gauge invariant observables. Unlike coordinate systems, which are simply labels on a manifold, reference frames are physical systems or sets of variables that represent material systems.
Their use in localising physical quantities directly addresses two interrelated challenges in GR: 
\begin{enumerate}
    \item[\textbf{(P1)}] The difficulty in constructing local gauge-invariant observables.
    \item[\textbf{(P2)}] The intention to provide a physical interpretation of gauge symmetries rather than interpreting them as mere mathematical redundancies or \enquote*{descriptive fluff} \citep{Earman2004-EARLSA}.\footnote{Problem \textbf{(P2)} is significant as gauge symmetries are pervasive across all known physical theories, including GR, warranting a physical interpretation of their ubiquity.}
\end{enumerate}

I will argue that these issues can be naturally addressed once physical quantities are localised using spatiotemporal reference frames rather than relying on manifold points or abstract coordinate labels.

To address the challenges outlined, I introduce a classification of reference frames in GR, conceptualised as sets of variables representing, or \textit{instantiated by}, material systems.\footnote{Throughout this paper, I will alternate between stating that a reference frame \textit{is} a physical system and that it is a set of variables in a mathematical model \textit{representing} a physical system. This distinction is conceptually important, and I believe the latter phrasing is more accurate. However, this nuance does not undermine the arguments presented herein. Thanks to Erik Curiel for his insightful suggestion on this matter.}  This classification exploits the fact that GR can be \textit{deparametrised} only for specific material models, enabling the construction of \textit{local} gauge-invariant Dirac observables.\footnote{Deparametrisation involves selecting a set of material variables $\{\phi\}$ that can serve as a spatiotemporal reference frame, at least locally. More rigorously, deparametrisation often requires rewriting Hamiltonian constraint in the form $H = \pi + h$, where $\pi$ are the conjugate momenta of $\phi$, and $h$, the \enquote*{physical Hamiltonian}, is independent of $(\phi, \pi)$. See \cite{Thiemann-k-essence,Tambornino2012}.} Agreed: using material reference frames is not the only viable approach. Early proposals, such as that championed by \cite{Komar}, explored the use of purely gravitational degrees of freedom to construct \textit{local} gauge-invariant Dirac observables, now known as \enquote*{Komar-Kretschmann scalars}. These are four scalar functions derived from the Riemann tensor, often (but ambiguously) referred to as \enquote*{intrinsic coordinates}. This clarifies that \lq{}reference frame\rq{} carries no intrinsic connotation of \lq{}material\rq{}; i.e. reference frames can be field-values that are not representative of matter.

In this work, I present three classes of material reference frames:
\textit{Idealised Reference Frames} (\textbf{IRFs})  represent systems where both the dynamical equations and the stress-energy contributions to the Einstein Field Equations (EFEs) are neglected.
\textit{Dynamical Reference Frames} (\textbf{DRFs}) incorporate dynamical equations but still neglect stress-energy contributions, as exemplified by systems analogous to \enquote*{test particles}.
Finally, \textit{Real Reference Frames} (\textbf{RRFs}) account for both stress-energy contributions and dynamical equations. Although \textbf{RRFs} are objects of great interest, as they are physically more realistic, this paper focusses on \textbf{IRFs} and \textbf{DRFs}. 

This classification provides fresh insights into the distinction between reference frames and coordinate systems and both \textbf{(P1)} and \textbf{(P2)}  above.
In particular, its significance lies in the way it contributes to these two considered substantive issues in GR. The taxonomy of reference frames is therefore intended not as a terminological contribution, but as a refined conceptual tool for articulating the relation between physical systems and mathematical structures.

The distinction between \enquote*{idealisation} and \enquote*{approximation} is also relevant to provide insights into the distinction between reference frames with coordinate systems.
According to \cite{Norton2012}, idealisation involves replacing the target system with a novel, often fictitious, system that simplifies the analysis, whereas approximations provide inexact descriptions \textit{of} the target system.
The difference lies in whether a novel system is introduced (idealisation) or not (approximation).
In GR, reference frames are best understood as structures within \textit{kinematically possible models} (KPMs), whose \textit{dynamically possible models} (DPMs) are subsets defined once dynamics is taken into account \citep{James_Read2023-mk}.
These models are represented as tuples $\langle \mathcal{M}, g_{ab}, \phi \rangle$, where $\mathcal{M}$ is a smooth manifold, $g_{ab}$ a Lorentzian metric, and $\phi$ the material degrees of freedom that may serve as spatiotemporal reference frames.\footnote{Abstract index notation \cite{Penrose1984} is used to emphasise the geometric nature of these objects, independent of specific coordinate representations.}
\textbf{IRFs} and \textbf{DRFs} emerge through successive approximations of the dynamics of the objects appearing in these models and assigned the role of reference frames.
In contrast, coordinates are idealisations lacking any physical instantiation.\footnote{Alternative perspectives on idealisations and approximations are discussed in \cite{frigg2022models}, where idealisations describe model-target relationships that must have physical interpretations, while approximations operate \lq{}solely at the mathematical level\rq{} and \lq{}involve no reference to a model\rq{} (ivi, p.318). Although beyond the scope of this work, it is worth exploring how the distinction between reference frames and coordinate systems aligns with such frameworks.} As such, they are not structures composing the tuples of general-relativistic models.

This paper aims to clarify the fundamental and ubiquitous concept of reference frame in physics.
The ubiquity of reference frames is encapsulated in the observations of several leading thinkers: Anderson notes that \lq{}\lq{}all measurements are comparisons between different physical systems\rq{}\rq{} \cite[p.128]{Anderson1967-en}; Rovelli reminds us that \lq{}\lq{}any measurement in physics is performed in a given reference system\rq{}\rq{} \cite{Rovelli1991}; and Landau and Lifshitz assert that \lq{}\lq{}for the description of processes taking place in nature, one must have a \emph{system of reference}\rq{}\rq{} \cite[p.1]{Landau1987-fh}.
These perspectives underscore the inadequacy of defining experimentally measurable quantities solely in terms of uninstantiated coordinates and highlight the importance of reference frames in capturing the phenomenology of physical processes \citep{BamontiGomes2024}.\footnote{For an analysis on the interplay between reference frames, phenomenology and ontology, see \cite{BamontiRelality}. For a defence of the \lq{}relationality of measuremnts\rq{}, see the \textit{unobservability thesis} in \cite{Wallace2022}.}

Moreover, a careful understanding of classical reference frames in GR provides an essential foundation for contemporary research on quantum reference frames \citep{RovelliQRF, Giacomini2019,Giacomini2021}. While the focus of this paper is on GR, the reference frame approach is also applicable to other gauge theories \citep{GomesRepresentationalscheme}.

In summary, this paper aims to clarify the fundamental concept of a reference frame in physics by analysing its role in overcoming the challenges of local gauge invariance and the proper interpretation of gauge symmetries in GR. In doing so, it draws a clear distinction between physically instantiated reference frames and abstract coordinate systems. This investigation serves not only as a conceptual \textit{memento} regarding the careful use of approximations in physical modeling \citep{Elgin2017-ry} but also as a necessary precursor to extending these ideas into the quantum domain.

\noindent The paper is structured as follows.

In Section \ref{sec2},  I explore the role of reference frames and coordinate systems within gravitational physics and review predominant definitions found in the literature.

Section \ref{sec3} provides a detailed classification of reference frames in GR, supported by concrete examples, and demonstrates how reference frames help resolve the issues of local gauge-invariant observables and the interpretation of diffeomorphism gauge symmetries.

Finally, Section~\ref{sec4} proposes an argument, based on modelling of measurement outcomes, for justifying the pragmatical overlap of \textbf{IRFs} and coordinate systems and discusses the implications of this overlap for our understanding of gravitational phenomena.

%In section \ref{sec5} we analyse the differences between \textbf{DRFs} and coordinates. We also highlight the usefulness of reference frames to address the problem of local gauge-invariant observables and to have a physical interpretation of diffeomorphism gauge symmetries in GR.

\section{Reference Frames vs. Coordinate Systems}\label{sec2}
\begin{quote}
    [Ehlers:] It is unfortunate that we use just one word. I think for science we need at least two different concepts, which are unfortunately denoted by the same word, namely, we use time in a first sense as a global parameter of events, to order them in a certain sequence, and that is not necessarily the same as what is measured by a good clock \cite[pp.234-6]{Barbour1995-BARMPF-7}.
\end{quote}
This section is not intended as a comprehensive review of all possible definitions of reference frames in spatiotemporal theories. Instead, its purpose is to contextualise my own definition and to extend the existing literature on the subject.
Additionally, I will outline the distinctions between coordinate systems and reference frames in gravitational and non-gravitational physics (\S \ref{sec2.1}).

Throughout this paper, I formally define a reference frame as a set of four scalar degrees of freedom of a physical system that establishes a local diffeomorphism $U \rightarrow V \subset \mathbb{R}^4$ for some $U \subset \mathcal{M}$ and $V \subset \mathbb{R}^4$. This mapping \textit{uniquely} assigning four numbers to each manifold point. For example, a set of four linearly independent scalar fields $\{\phi^{(I)}\}_{I=1,\dots,4}$ satisfying dynamical equations (e.g., Klein-Gordon equations) can serve as a reference frame.\footnote{Besides the aforementioned Komar-Kretschmann scalars, the usage of scalar fields is consistent with, and conceptually supported by much of the extant  literature. For example, Hardy notably employed scalar fields to define reference systems---though often without using my terminology \citep{Hardy:2016snq}. See also \cite{Westman:2007yx}.}
The metric field can then be expressed relationally as $g_{IJ}(\phi):= \big[ (\phi^{(I)})^{-1} \big]^* g_{ab}$, where the symbol \([\bullet]^*\) denotes the pullback of the metric via the scalar fields.\footnote{It is very difficult to think of a realistic situation in which a reference frame would cover the entire manifold. In fact, four Klein-Gordon scalars could end up having the same values everywhere on $\mathbb{R}^4$, thus representing only one (physical) point. Thus, in order to indicate viable reference frames, $\{\phi^{(I)}\}$ should be \textit{at least} \textit{locally} invertible, i.e. in some open set $U \subset \mathcal{M}$ and for a given chart, the Jacobian det$(\partial \phi^{(I)}/\partial x^\mu)\neq 0 $.\label{jacobian}}
This relational framework treats reference frames as physical degrees of freedom on par with other systems and provides effective means to address the problems \textbf{(P1)} and \textbf{(P2)} discussed earlier. As argued in this paper, such relational approaches offer distinct advantages over more traditional definitions in the foundations of GR.

Let me now summarise some of the influential definitions of \lq{}reference frames\rq{} that have shaped the literature.

Famously, the work of Norton and Earman defines a reference frame in terms of a smooth, timelike \textit{4-velocity} field $U^a$ tangent to the worldlines of a material system, to which an equivalence class of coordinates is locally adapted \citep{Earman1973, Earman1974, Earman1978}. This definition is widely adopted in GR literature (see the more recent \cite[p.1042]{Bradley2021}, or \cite[p.4]{JacobsForthcoming-JACHNT-2}).
Given its prominence, I quote at length the definitions of a reference frame provided by Earman and Norton:

\begin{quote}
In this context a reference frame is defined by a smooth, timelike vector field $V$. [...]Alternatively, a frame can, at least locally, be construed
as an equivalence class of coordinate systems. The coordinate system $\{x^i\}$,
$i = 1, 2, 3, 4$, is said to be adapted to the frame $F$ if the trajectories of the vector field which defines $F$ have the form $x^a = \text{const}, a = 1, 2, 3$. [\dots] $F$ may be identified with a maximal class of internally related class of coordinate
systems. \cite[p.270]{Earman1974}
\end{quote}

\begin{quote}
    [...] it is now customary to represent the intuitive notion of a physical frame of reference as a congruence of time-like curves. Each curve represents the world line of a reference point of the frame. 
    [...] A coordinate system $\{x^i\}$ $(i = 1, 2, 3, 4)$ is said to be ‘adapted’ to a given frame of reference just in case the curves of constant $x^1$, $x^2$ and $x^3$ are the curves of the frame. These three coordinates are ‘spatial’ coordinates and the $x^4$ coordinate a ‘time’ coordinate. \cite[p.209]{Norton1985}

    Thus a frame of reference is introduced in standard practice as a congruence of timelike curves defined on the manifold (with metric). The frame, if smooth, assigns a velocity, its tangent vector, to every event in the manifold. \cite[p.1242]{Norton1989}
\end{quote}

I will examine this approach in detail in \S \ref{sec3.2.3}, by relating it to my classification. However, it is essential to establish immediately that the quadruple of scalar fields 
$\{\phi^{(I)}\}$, while sufficient to define a local diffeomorphism 
$U \subset \mathcal{M} \to V \subset \mathbb{R}^4$, does \textit{not} constitute in general the  components of a spacetime 4-velocity. 
Each $\phi^{(I)}$ transforms as a scalar under diffeomorphisms, and the ordered quadruple 
$(\phi^{(0)},\phi^{(1)},\phi^{(2)},\phi^{(3)})$ merely provides a system of 
relational labels for points. 
A genuine 4-velocity field $U^a$, by contrast, 
is a vector field in $T_p\mathcal{M}$, subject to normalisation 
($U^a U_a=-1$) and possessing a clear dynamical interpretation as the tangent  to timelike worldlines. This distinction will become central when comparing  the present framework with the orthodox view in §\ref{sec3.2.2}.

\noindent Closely related to these definitions is the characterisation of reference frames using \emph{tetrads} (or \enquote*{orthonormal frames}) \cite[ch.3.4]{Wald1984}; see also \cite{Duerr2021}. Tetrads \(e^a_{(I)}\) are four smooth vector fields satisfying the orthonormality condition $e^a_{(I)} e^b_{(J)} g_{ab} = \eta_{IJ}$, with $\eta_{IJ} = \text{diag}(-1, 1, 1, 1)$. 
The indices $I,J=\{0,\dots,3\}$, called \textit{internal Minkowski indices}, label the tetrads forming an \textit{orthonormal basis} for the tangent space  $T_p\mathcal{M}$, effectively providing a local map to Minkowski space.\footnote{Strictly speaking, the tangent space \( T_p(\mathcal{M}) \) is already Minkowskian, so tetrads do not map it to a separate Minkowski space but establish a local orthonormal basis within \( T_p(\mathcal{M}) \) itself.}
This allows tensor fields to be locally expressed in terms of \textit{tetrad components}, simplifying their local form. For instance, the metric becomes $g_{ab} = e^{(I)}_a e^{(J)}_b \eta_{IJ}$, and a vector field $U^a$ takes the form $U^a = u^{(I)} e^{a}_{(I)}, \quad \text{with } (u^{(0)}, \dots, u^{(3)}) \in \mathbb{R}^4.$ 
Also, each point \( p \in \mathcal{M} \) can be associated with four independent scalar values \( X^{(I)} \) instead of the usual four coordinates \( x^\mu \). 
In essence, each tetrad is a 4-vector representing a \textit{direction} in spacetime: the temporal tetrad \( e^a_{(0)} \) defines the local temporal direction, often associated with the 4-velocity of a \textit{comoving} observer, while the spatial tetrads \( e^a_{(1)}, e^a_{(2)}, e^a_{(3)} \) define the local spatial directions.
Together, they form a \enquote*{\textit{tetrad frame}}.

Recent works have also sought to distinguish reference frames from coordinate systems by linking reference frames to an observer's state of motion.
For example, \cite[\S4.3]{pooley2022reality} defines a reference frame as a set of standards (e.g., rest and simultaneity) relative to which motion can be quantified (see also \citealp{sep-spacetime-iframes}). This paper demonstrates that such definitions are valid, but insufficient to exhaustively characterise reference frames in GR.

Similarly to what I do here, in quantum contexts, reference frames are tied to physical material systems, ultimately quantum in nature \citep{CastroRuiz2020}. Neglecting this quantum nature by treating reference frames merely as coordinates overlooks essential properties. However, many studies on quantum reference frames \textit{implicitly} assume non-backreacting material systems \citep{delaHamette2023, QuantumHole, Geng}.

%Ambiguities between reference frames and coordinate systems still persist, particularly in classical GR. For instance, \cite[p.215]{Read2020} conflates reference frames and coordinate systems by defining \enquote*{non-tensorial objects} as frame-dependent objects and then defining a tensorial object as a \lq{}coordinate-dependent\rq{} object (ivi, p.217). Similarly, \cite{LEHMKUHL2014} uses these terms interchangeably, reflecting Einstein's original practice.
%These cases illustrate that, even if the conflation between coordinates and reference frames does not necessarily undermine the validity of the work, it is crucial, in my view, to explicitly acknowledge the distinction, especially when addressing issues of measurability and observability.

I now turn to the challenging task of comparing the role of reference frames in GR with their role in pre-GR theories.

%A local coordinate system in a N-dimensional topological manifold is unanimously considered to be a choice of a local \textit{chart}, \textit{i.e.} an open set and a homeomorphism $\gamma$ which assign N labels to a point of the manifold. Formally, a local coordinate system (also referred to as \enquote*{coordinate chart}) is defined as a $1:1$, onto map $\gamma:\mathcal{S}\rightarrow R^N$ from an open patch $\mathcal{S}\subset M$ of the manifold $M$ into the N-fold product of the real numbers.

\subsection{GR vs. Pre-GR Physics}\label{sec2.1}

To understand the nuanced role of reference frames and coordinate systems in GR, it is instructive to compare their meaning in GR with their counterparts in what I call \enquote*{pre-GR physics}. This comparison elucidates the conceptual distinctions and practical implications of these two constructs in different physical frameworks.
In particular, I will focus on the experimental role of material reference frames, investigating whether and how measurements of a quantity are influenced by the reference frame and under what conditions this influence becomes negligible.

To facilitate this discussion, I will frequently employ the term \enquote*{\textit{dynamical (un)coupling}}.  In general, two fields (for instance, a scalar field \(\phi\) and a metric field \(g_{ab}\)) are said to be \textit{dynamically coupled} if the dynamics of one field \textit{influences} the dynamics of the other. This also applies \textit{if this influence is not necessarily reciprocal}: for example for non-backreacting fields. Conversely, I consider two fields as \textit{dynamically uncoupled}, if any influence is absent.\footnote{An alternative perspective on dynamical coupling, as presented in \cite{BamontiGomes2024}, understands coupling in terms of what \cite{BamontiCOUPL} calls \textit{correlation}. Two fields can exhibit correlation even in the absence of mutual influence. From this broader perspective, any field in a spatiotemporal theory is correlated, at least indirectly, to every other field through the metric. It is evident that the only way to uncorrelate a field \textit{from a metric field} in a generic spatiotemporal theory is to disregard its dynamical equations, since in \textit{any} spatiotemporal theory the dynamics is always written relative to a metric. In this work, I decide not to follow \cite{Lehmkuhl2011-LEHMOT}'s distinction between influence and coupling.
For my purpose, I simply focus on the notion of coupling in terms of influence between fields.
However, it remains true that everything that I will say does not in any way contradict the works cited above in this footnote.}

%In what follows, I will argue how fields can be uncoupled and the differences in this procedure between GR and pre-GR theories, such as Special Relativity and Newtonian mechanics.

%Conversely, uncoupling is a stronger condition: I define two fields to be uncoupled if the dynamics of each does not depend on the other.
%It is evident that the only way to uncouple a field \textit{from a metric field} in a spatiotemporal theory is to disregard its dynamical equations, since in \textit{any} spatiotemporal theory the dynamics is always written relative to a metric.

To begin with, I will adopt the following \textit{coarse-grained} characterisation:\footnote{I will omit that, according to their definition, both provide a local diffeomorphism $U \subset \mathcal{M} \rightarrow V \subset \mathbb{R}^4$, for some $U \subset \mathcal{M}$, which \textit{uniquely} assigns four numbers to each point  $p \in U$. This means, for example, that not any set of dynamically coupled degrees of freedom constitute a reference frame (see also fn.\ref{jacobian}).}
\begin{itemize}
    \item \textbf{Reference Frame:} A set of dynamically coupled and \emph{instantiated} physical degrees of freedom.
    \item \textbf{Coordinate System:} A set of non-dynamical and \emph{uninstantiated} mathematical labels.
\end{itemize}

Here, the dynamical coupling of the reference frame is understood with respect to the \emph{dynamical system of interest}—in GR, the metric field. The term \enquote*{instantiated} implies that these degrees of freedom correspond to physical, or at least physically possible, objects.

\begin{comment}
The logical relationship between these two notions can be summarised as follows:
\[
\text{Dynamically Coupled} \implies \text{Instantiated}.
\]
That is, any field dynamically coupled to another necessarily has dynamics and, by definition, represent a physical or physically realisable entity.
\end{comment}

\subsubsection{Reference Frames and Coordinates in Pre-GR Physics}

In pre-GR theories, the distinction between reference frames and coordinate systems is \textit{conceptually} significant but less pressing in experimental practice.

In many cases, a reference frame can be identified with an instantiated coordinate system without resorting to any approximation procedures. So, its role in the experimental practice is often irrelevant from the outset.

For example, consider Maxwellian electrodynamics in Minkowski spacetime. Here, the electromagnetic field—the 'field of interest'—is treated as a subsystem of the universe that does not affect the global inertial reference frame, encoded by the Minkowski metric \( \eta_{\mu\nu} = \text{diag}(-1,1,1,1) \).
This inertial reference frame may be instantiated by non-electrically charged rods and clocks, which have no influence relationship with the electromagnetic field. In other words, the dynamics of the electromagnetic field is completely unaffected from the reference frame, so its distribution and propagation remain unchanged regardless of the physical instruments used.

As Einstein stated \cite[p.38]{Einstein1905}:
\begin{quote}
    The theory to be developed—like every other electrodynamics—is based upon the kinematics of rigid bodies, since the assertions of any such theory concern relations between rigid bodies (systems of coordinates), clocks, and electromagnetic processes.
\end{quote}
This insight shows that reference frames in pre-GR can function as \enquote*{instantiated coordinate systems}: they are, by definition, dynamically uncoupled (in the strict sense of not sharing influence relations) from the field of interest while still being  instantiated by physical objects.\footnote{From now on, I will not repeat that I understand dynamical uncoupling in terms of ‘non-influence’.}
This intermediate notion—instantiated coordinate systems—acts as a conceptual bridge between the abstract concept of coordinates and the physically instantiated reference frames.\footnote{As I will show in the next section, this is also the role of \textbf{IRFs}. However, contrary to the case of special relativity, in GR \textbf{IRFs} emerge only as the result of an approximation procedure.}

Although there is a practical equivalence between reference frames and coordinate systems in many pre-GR contexts, the conceptual distinction remains. In general-covariant formulations of Maxwell electrodynamics in Special Relativity, for example, coordinates remain mere uninstantiated parameters devoid of physical meaning, just as in GR.
Thus, although reference frames may coincide with coordinate systems in practice, this equivalence is not fundamental but rather contingent on the experimental context.

In light of this discussion, I propose the following
\textbf{refined charachetrisation (pre-GR):}
\begin{description}
    \item[\textbf{Reference Frame:}] A set of dynamically coupled and \textit{instantiated} physical degrees of freedom.\footnote{For instance, consider using four complex scalar fields as a reference frame for an electrodynamical system in Minkowski spacetime. In this case, the electromagnetic field (the \textit{dynamical quantity of interest}) and the scalar fields are coupled. These scalar fields act as sources for the electromagnetic field and are themselves influenced by it. The presence of this coupling can be seen in the standard \( U(1) \) Lagrangian density:
    \[
    \mathcal{L} = \left[(\partial_a \phi \partial^a \phi^* - m^2 \phi \phi^*) - \frac{1}{4} F^{ab} F_{ab}\right] 
    + e^2 \left[A_a A^a \phi \phi^* + \frac{1}{e} A^a J_a\right],
    \]
    where the first terms describe the free field dynamics of the complex scalar field and the Maxwell field, and the last terms represent their interaction. Here, \( e \) is the electric charge, \( m \) is the mass of the scalar field, and \( J^a = i e[-\phi \partial^a \phi^* + \phi^* \partial^a \phi] \) is the conserved current. For a more intuitive example, think of \textit{electrically charged} rods and clocks in \cite{Einstein1905}'s proposal for an electrodynamic theory in Minkowski spacetime.\label{fnlagrangian} Finally note that, of course, a standard spacetime reference frame needs one “time” function and three “space” functions. Since the four complex fields give \textit{eight} real degrees of freedom in total, we have to extract four independent, $U(1)$-invariant, real-valued functions that define a local reference frame.} 

    \item[\textbf{Reference Frame/Coordinate System (interchangeable):}] A set of \enquote*{definitionally} dynamically uncoupled and \textit{instantiated} labels, where no approximation procedure is required to uncouple the frame from the dynamical system under study.

    \item[\textbf{Coordinate System:}] A set of non-dynamical and \textit{uninstantiated} mathematical labels, typically used in the context of general covariant formulations where the coordinates lack physical instantiation.
\end{description}

It is important to distinguish between being \enquote*{dynamically uncoupled} and being \enquote*{non-dynamical.} Dynamically uncoupled objects are instantiated and possess their own dynamics, but this dynamics is neglected. In contrast, non-dynamical objects are not instantiated, meaning that no dynamics can be attributed to the variables constituting the coordinate system.\footnote{At most, gauge conditions can be imposed on a coordinate system.} 

I now turn to how these notions must be refined in GR.

\subsubsection{Reference Frames and Coordinates in GR}

In GR the situation is fundamentally different because there is no \lq{}gravitationally neutral\rq{} physical system. In this context, reference frames cannot be dynamically uncoupled without adopting approximations, as no physical system is entirely free from the influence of gravity.\footnote{The reference frame might be considered dynamically uncoupled if the interaction effects are negligible compared to the \textit{experimental} precision. Nevertheless, in this context, the property of being dynamically uncoupled is independent of experimental errors and constraints.} 

When I refer to GR, I mean the dynamical theory \emph{of} the metric field, which is the dynamical field of interest. Nevertheless, the discussion applies equally well to any other field within a general-relativistic framework.

Consider again the example of Maxwellian electrodynamics.

In Special Relativity, the Minkowski metric is fixed and remains unaffected by the electromagnetic field. Moreover, the electromagnetic field is uncoupled from the (neutral) measuring instruments that instantiate the reference frame; although he dynamics of both depend on the Minkowski metric, the dynamics of one do not influence the other.
In particular, the value of the electric or magnetic field \textit{at a point} is unaffected from the presence of the (neutral) instruments/frames used to assign the location.

In contrast, in GR the metric is dynamical and is influenced by both the electromagnetic field and the measuring instruments. One must account for the mutual influence between the gravitational field and the degrees of freedom defining the frame.
This \textit{mutual} interaction, encoded in the EFEs containing the stress-energy contents of the material frame,  prevents us from identifying reference frames with mere instantiated but uncoupled coordinate systems, as is often feasible in pre-GR physics.
In GR reference frames  (i.e. measuring instruments such as rulers and clocks) are intrinsically coupled to gravity and, \textit{indirectly}, to the electromagnetic field (\textit{even if the reference frame is neutral with respect to electromagnetic charge}).
\begin{comment}
\footnote{To further illustrate, consider the general-relativistic analogue of the example in footnote~\ref{fnlagrangian}. So, we have GR coupled with a Maxwell field and a charged scalar field. In this case, the total Lagrangian (density) is given by:
\[
\mathcal{L} = \left( g^{ab} D_a \phi D_b \phi^* - m^2 \phi \phi^* \right) - \frac{1}{4} g^{ab} g^{cd} F_{ac} F_{bd} - R,
\]
where 
\[
D_a \phi = \nabla_a \phi + i e A_a \phi
\]
is the covariant gauge derivative, \(\nabla_a\) represents the curved connection, and \(R\) denotes the Ricci scalar constructed from the metric \(g_{ab}\).
A notable feature of this formulation is the absence of free field Lagrangian terms: gravity \lq{}\textit{sticks}\rq{} to the other fields through the covariant derivative, intertwining their dynamics with its own. This coupling encapsulates that in GR all fields are inherently coupled with gravity and \textit{via} the universalityof gravitational interaction.}
\end{comment}

In fact, the electromagnetic field is coupled to the gravitational field because the metric governs the propagation of electromagnetic fields through Maxwell's equations in curved spacetime, while the electromagnetic field, in turn, influences the gravitational field through its stress-energy content.
Thus, even if a reference frame is neutral, it is indirectly coupled to the electromagnetic field through their mutual interaction with the common gravitational field.
If the electromagnetic field is intense, it can significantly influence the metric and, consequently, alter the way the reference frame instruments measure spatial and temporal location.

This discussion underscores the conceptual distinction between reference frames and coordinate systems in GR. Coordinate systems in GR are purely mathematical artefacts without physical instantiation. If a coordinate system were instantiated, it would necessarily interact gravitationally, thereby becoming a set of dynamically coupled degrees of freedom—that is, a genuine reference frame.
Thus, in GR the notions of reference frame and coordinate system cannot coincide either conceptually or in experimental practice.

The differences between GR and pre-GR physics highlight the centrality of this distinction. While pre-GR theories sometimes allow for an overlap between reference frames and coordinate systems under specific conditions, GR demands a clear separation because all physical objects that can instantiate a reference frame inevitably influence and are influenced by gravity.

In \S \ref{sec3}, I will explore how approximations in GR enable the practical use of dynamically (un)coupled reference frames, while maintaining their conceptual distinction from coordinate systems.
This analysis builds on and extends the themes introduced in previous sections, providing a robust framework to address the challenges posed by problems \textbf{(P1)} and \textbf{(P2)}.

\subsubsection{The Newtonian case}

Newtonian gravity must be included under the umbrella of \lq{}pre-GR physics\rq{}. However, it is important to recognize that Newtonian theory is still fundamentally gravitational. Therefore, the terms \lq{}pre-GR\rq{} and \lq{}non-gravitational\rq{} should not be used interchangeably. 
In what follows I will focus \textit{exclusively} on the standard Newtonian theory as formulated with absolute space and absolute time. However, it is important to acknowledge that several alternative Newtonian framework exist---for example, Galilean spacetime, where absolute space is replaced by a flat affine connection and absolute simultaneity is preserved; and Newton–Cartan theory, which arises from gauging the Bargmann group and is structurally closer to GR (see \citealp{Hartong2023}, \citealp{BamontiBI}). A full comparative analysis would exceed the scope of this section, but some relevant remarks will be offered below. 

In both Newtonian gravity and GR material systems that instantiate reference frames are not gravitationally neutral and cannot be overlooked in experimental practice.
In Newtonian gravity, every mass both generates a gravitational field and experiences its pull. Likewise, in GR \textit{all forms of energy} source—and are influenced by—the gravitational field (see \cite{Brown2013} for an in-depth analysis of the principle of action and reaction).\footnote{Please note: action-reaction in Newton is \textit{not} about spacetime as it is in GR.}

As a result, both Newtonian and relativistic frameworks challenge the notion of reference frames that are wholly uncoupled from gravity. Achieving such decoupling \textit{always requires approximation procedures.}

One way to disentangle these ideas is via Pooley’s distinction between different coordinate interpretations in pre-GR contexts \cite[\S 8.10]{pooley2022reality}. Pooley identifies two key interpretations:

\paragraph{The ESR (\citeauthor{Einstein1905}–\citeauthor{Stachel1993}–\citeauthor{Rovelli2004}) Interpretation:}
In this view, coordinates are \textit{anchored} to material objects—such as synchronised clocks and rods—which establish measurable time and distance intervals.
Under this view, coordinates can be instantiated by material objects whcih are dynamically uncoupled from the system under study (recall the Maxwellian example discussed earlier). Rovelli eloquently summarises this point in the Newtonian case:
\begin{quote}
    For Newton, the coordinates \(\vec{x}\) that enter his main equation
    \[
    \vec{F} = m \frac{d^2 \vec{x}(t)}{dt^2}
    \]
    are the coordinates of absolute space. However, since we cannot directly observe space, the only way we can coordinatise space points is by using physical objects. The coordinates \(\vec{x}\) [...] are therefore defined as distances from a chosen system \(O\) of objects, which we call a \enquote*{reference frame.} [...] Notice also that for this construction to work, it is important that the objects \(O\) forming the reference frame are not affected by the motion of the object \(A.\) \textit{There shouldn’t be any dynamical interaction between} \(A\) \textit{and} \(O.\) \cite[pp.61–62]{Rovelli2004} (emphasis added)
\end{quote}

However, due to the universality of gravitation, achieving this lack of interaction requires approximation procedures. Thus, it follows that in the ESR view:
\begin{itemize}
    \item In gravitational physics (both GR and Newtonian gravity), reference frames are dynamically uncoupled from the system under study only through approximations; coordinates remain, by definition, non-dynamical and uninstantiated.
    \item In non-gravitational physics, reference frames can be dynamically uncoupled without requiring approximations (as in the Maxwellian example), allowing them to be used interchangeably with instantiated coordinates.
\end{itemize}

\paragraph{The ATF (\citeauthor{Anderson1967-en}–\citeauthor{Trautman1967}–\citeauthor{Friedman2016-br})  Interpretation:}
In this view, coordinates are not anchored to external material objects but are instead defined via gauge-fixing conditions that \emph{reveal} physically meaningful \textit{spacetime structures}. 
For Newtonian physics in its absolute space–absolute time formulation, these  structures are the temporal metric defining a universal time flow, and the Euclidean spatial metric defining an absolute three-dimensional arena. Gauge-fixing conditions identify these background structures by selecting a privileged global time coordinate aligned with absolute time, and a rigid system of spatial coordinates aligned with absolute space. 
So, in the ATF interpretation, coordinates are instantiated by the spatio-temporal background  structure rather than by material objects.
This interpretation aligns with a substantivalist interpretation of spacetime, according to which, very broadly, space-time exists as a \textit{sui generis} kind of \lq{}substance\rq{}, independent of material content (see \cite{Brown2013} for an analysis on the category of \lq{}substance\rq{}).
Coordinates \textit{encode} spatiotemporal physical magnitudes and are still anchored to the real world, but this anchoring does not depend on external material objects, it is intrinsic to the spatiotemporal structure of the theory.

The debate between these interpretations is central to our understanding of absolute space and time coordinates in Newtonian physics and to clarifying the relationship between reference frames and coordinates. This leads to a deeper ontological question: Are absolute space and time part of the physical inventory of the universe, or do they belong to a \textit{meta}-physical category? The answer to this question has significant implications for how we interpret reference frames and coordinates.

Before turning to this ontological issue, it is worth adding a brief complementary note. Alternative Newtonian frameworks reshape the terms of the debate. In Galilean spacetime, where absolute space is absent, tthe ATF interpretation loses its force, since there is no absolute space background structure available to instantiate coordinates.\footnote{For Galilean physics, whose KPMs are given by tuples $\langle \mathbb{R}^4,\, t^a,\, h_{ab},\, \overset{0}{\nabla^a}_{bc} \rangle$, these gauge-fixing conditions include imposing:
\[
\overset{0}{\Gamma^\mu}_{\nu\rho} = 0, \quad t^\mu = (1, 0, 0, 0), \quad h_{\mu\nu} = \text{diag}(0, 1, 1, 1),
\]
where \(t^\mu\) and \(h_{\mu\nu}\) represent the temporal and spatial metrics, respectively, and \(\overset{0}{\Gamma^\mu}_{\nu\rho} = 0\) represents the flat connection (that is, the components of $\overset{0}{\nabla^a}_{bc}$).
 Absolute space is eliminated in favour of an even more abstract entity, the (non-unique) flat-affine connection.} 
 By contrast, Newton–Cartan theory should be regarded as conceptually much closer to relativistic gravitation than to a pre-GR setting. A detailed investigation of these variants would require a separate treatment; here I have confined myself to Newton’s standard absolute space–absolute time framework.

\paragraph{Newtonian Space and Time: Two Ontological Cases}

\cite{newton1687principia} defines absolute space and time as:
\begin{itemize}
    \item \textbf{Absolute Space:} \enquote{Absolute space, in its own nature, without regard to anything external, remains always similar and immovable.}
    \item \textbf{Absolute Time:} \enquote{Absolute, true, and mathematical time, of itself, and from its own nature flows equably without regard to anything external.}
\end{itemize}

In Newton’s framework, absolute space and time serve as the stage upon which physical phenomena unfold, yet they do not interact causally with physical bodies or with our senses. Even if considered as substances, they are \textit{sui generis}: causally inefficacious and unaffected \cite[p.4]{Brown2013}.
This dual status—being \lq{}real\rq{} yet causally inert—renders them unique in ontological debates.

In particular, this status leads to two possible ontological interpretations of Newtonian space and time, contingent upon whether they are \lq{}part of the ontology\rq{} or not:\footnote{By asserting that absolute space and time are \enquote*{part of the ontology,} I mean that they are included in the fundamental inventory of what exists in the physical world.
In contrast, if absolute space and time are \emph{not} part of the ontology,\textit{ they still exist} but are understood as \textit{meta}-physical entities. This means they transcend the physical domain and belong to a different category of existence, common to e.g. the existence of numbers, concepts, ideas etc.}

\begin{description}

\item{\textbf{Case 1: Absolute Space and Time are Part of The Physical Ontology.}}
One ontological possibility is that absolute space and time are part of the physical inventory of the universe
As such, they play an indispensable role in describing physical phenomena and are considered to have an objective, \textit{albeit causally inert}, role within the physical realm.
In this case, the ATF approach remains viable because coordinates are \textit{instantiated} by the background absolute space and time. 

However, in \cite{newton1999principia}’s \textit{General Scholium}, absolute space is also described as the \enquote{\textit{sensorium Dei}} (God’s sensorium) —a divine medium through which God perceives the universe. Consequently, according to this reading, absolute space and time are \lq{}real\rq{} but not physical; they are \textit{meta}-physical frameworks underpinning the physical world rather than being entities within it.
This complicates the ATF interpretation of coordinates, as it frames space as a divine manifestation rather than a conventional physical substance.

\item{\textbf{Case 2: Absolute Space and Time Are Not Part of The Ontology.}}

If absolute space and time are not part of the ontology, then they cannot \textit{instantiate} coordinates.
In this case, the substantivalist view collapses, invalidating the ATF interpretation. Thus, the ESR approach becomes the \textit{only} viable framework: in Newtonian gravity, spatiotemporal reference frames must be instantiated by material objects, which then require approximation procedures to \enquote{screen out} gravitational interactions, effectively relegating them to dynamically uncoupled labels.
\end{description}

In the forthcoming section, I will introduce my novel three-fold classification scheme for reference frames in GR, which elucidates the requisite approximation procedures essential for developing uncoupled reference frames. This classification reflects how the reference frame is coupled with the gravitational field.

\section{IRF, DRF, RRF} \label{sec3}

Building on the distinctions developed in \S \ref{sec2}, I now explore the different possibilities to define a reference frame as a set of four independent scalar degrees of freedom, providing a relational localisation of spacetime points. 
In particular, I will introduce the three types of reference frame in GR:

\begin{enumerate}
    \item \textbf{Idealised Reference Frames (IRFs),} which neglect both the stress–energy and dynamical equations of the material system;
    \item \textbf{Dynamical Reference Frames (DRFs),} which include the frame’s equations of motion but still ignore its stress–energy backreaction;
    \item \textbf{Real Reference Frames (RRFs),} which fully couple the matter’s dynamics and stress–energy to the Einstein equations.
\end{enumerate}

\subsection{Idealised Reference Frames (IRFs)} \label{sec3.1}

An Idealised Reference Frame (\textbf{IRF}) is defined by neglecting any dynamical interaction of the material system that constitutes the reference frame. Specifically, two approximations are adopted:
\begin{enumerate}
    \item[(a)] The contribution of the material reference frame to the stress-energy tensor in the EFEs is neglected. \label{approxstress}
    \item[(b)] The equations governing the dynamics of the material reference frame itself are ignored. \label{approxdyn}
\end{enumerate}

Significantly, ignoring the dynamical equations of the reference frame implies treating the reference frame as non-gravitating, thereby leading to an apparent underdetermination, as I will demonstrate below.

My \textbf{IRFs} closely resemble what \cite{Rovelli2004} (p.62) refers to as \enquote*{undetermined physical coordinates.} As Rovelli explains:
\begin{quote}
    We obtain a system of equations for the gravitational field and other matter, expressed in terms of coordinates \(X^\mu\) that are interpreted as the spacetime locations of reference objects whose dynamics we have chosen to ignore. This set of equations is underdetermined: the same initial conditions can evolve into different solutions. However, the interpretation of such underdetermination is simply that we have chosen to neglect part of the equations of motion.
\end{quote}

Although Rovelli’s terminology provides a useful analogy, I refrain from adopting it here in order to avoid unnecessary confusion between the concepts of \enquote*{reference frame} and \enquote*{coordinate system.} Moreover, I contend that the term \enquote{indeterminism of the dynamics} is the most suitable in this context, given that \textit{underdetermination} usually pertains to a surplus of possible options that are not connected to dynamical aspects.

As hinted above, approximation \eqref{approxdyn} introduces a form of indeterminism in the dynamics of the metric field when expressed in the frame of the matter degrees of freedom.
To illustrate this, consider a metric field \(g_{ab}\) satisfying the EFEs and four scalar fields \(\{\phi^{(I)}\}\) representing the \textbf{IRF}.
Via the pullback, one constructs the relational observable $g_{IJ}(\phi):=\big[ (\phi^{(I)})^{-1} \big]^* g_{ab}$ , representing a configuration \((g_{ab}, \phi^{(I)})\) of the fields assuming such and such values.

The set of four scalar fields serves as a local reference frame akin to one clock and three rods.\footnote{Arguably, the scalar field selected to play the role of the timelike variable (say $\phi^{(1)}$) needs to satisfy some properties such as a homogeneity condition $\nabla^i\nabla_i\phi^{(1)}(x^\mu)=0$ where $i=1,2,3$ are spatial indices in some coordinates $\{x^\mu\}$. We could also assume a \lq{}monotonicity condition\rq{} connected with some assumptions on its potential (when it is considered).}
The relational observable \(g_{IJ}(\phi)\) emphasises that the relational metric is not localised on the points of the manifold \(\mathcal{M}\) but rather on the space of ordered four-tuples of scalar field values, \(\mathbb{R}^4\).
In this sense, the scalar fields can be viewed as diffeomorphisms mapping \(g_{ab}\) on \(\mathcal{M}\) to \(g_{IJ}\) on \(\mathbb{R}^4\), thereby encoding spatiotemporal localisation in relational terms—a viewpoint consistent with Einstein's notion of \enquote{coincidences} \citep{Einstein1916} and elaborated in \cite{Giovanelli2021}.\footnote{\textit{Relational localisation} implies that a spacetime point \(p\) is identified through the inverse relation:
\[
p = (\phi^{(I)})^{-1}(x), \quad x\in \mathbb{R}^4 :=\{\text{the point where the scalar fields take a specific set of values}\}.
\]
Thus, spatiotemporal localisation is expressed in terms of the matter fields serving as the reference frame.
Such perspective is discussed extensively in the literature on relational observables and diffeomorphism invariance \citep{Rovelli2004, Goeller2022}.}

The apparent indeterminism arises because both the configuration \((g_{ab}, \phi^{(I)})\) and the configuration \(([d^*g]_{ab}, \phi^{(I)})\), for any diffeomorphism \(d \in \text{Diff}(M)\), are equally valid representations of the dynamics.

This redundancy is a direct consequence of the dynamical uncoupling of \textbf{IRFs} from the metric field.
As a result, the metric evolution is determined only up to four arbitrary functions.
Importantly, this is not a pernicious indeterminism but rather the manifestation of an unexpressed gauge freedom in the dynamics; the same initial data can evolve into different solutions that are gauge-related. This kind of indeterminism is also evident when employing coordinates in GR (see, for example, the discussion of the \emph{hole argument} in \citealp{Earman1987-EARWPS,Weatherall2018-WEARTH-2,HOLE}). As noted in \cite{BamontiGomes2024}, the relational observable \(g_{IJ}(\phi)\) is \emph{not} gauge-invariant if the set \(\{\phi^{(I)}\}\) forms an \textbf{IRF}.

A nuanced discussion of the intersection between \textbf{IRFs} and coordinate systems is reserved for \S\ref{sec4}. In essence, \textbf{IRFs} may be viewed as \enquote*{instantiated coordinate systems}. However, analysing this overlap helps shed light on the difference between reference frames and coordinates.

\subsubsection*{A Remark on Approximation (b) and Its Implications}

One might consider the possibility of adopting only approximation (b), thereby allowing the material reference frame to contribute to spacetime curvature while neglecting its own dynamical equations. However, this scenario still leads to indeterminism. The approximate nature of this approach becomes immediately evident from the Bianchi identities:
\begin{equation}
\nabla_a G^{ab} = 0,
\end{equation}
which, via the EFEs, imply \(\nabla_a T^{ab} = 0.\)

In GR, the Euler-Lagrange equations for the matter fields are essentially equivalent to enforcing \(\nabla^a T_{ab} = 0.\) 
This equivalence illustrates why the EFEs are said to \textit{explain} the motion of matter \citep{Weatherall2016-WEAIME}.
Consequently, allowing a non-vanishing \(T_{ab}\)  that satisfies \(\nabla^a T_{ab} = 0\), but without explicitly considering the matter equations of motion clearly constitutes an approximation.

Although theoretically possible, to the best of my knowledge, no concrete examples of such approximations exist in theoretical practice. For this reason, I choose not to consider this possibility as a distinct class of reference frames in this work.

\subsection{Dynamical Reference Frames (DRFs)}\label{sec3.2}

In contrast to Idealised Reference Frames (\textbf{IRFs}), which neglect both the stress-energy contribution and the dynamical equations of the material system, a Dynamical Reference Frame (\textbf{DRF}) is obtained by adopting only approximation (a) above. One neglects the stress-energy contribution of the material system to the EFEs while retaining its dynamical equations. Thus, the dynamical equations of the material reference frame effectively serve as gauge-fixing conditions, thereby eliminating gauge redundancy and yielding a fully deterministic evolution.

This perspective establishes a key correspondence: once the dynamical equations of the reference frame are specified, the parameters of the reference frame essentially become equivalent to coordinates within a gauge-theoretic framework. The imposed gauge conditions fix the redundancy, ensuring that the evolution of the system is deterministic. In effect, using coordinates alongside gauge freedom is conceptually analogous to employing a dynamical reference frame—one instantiated by a physical system that is dynamically coupled to the gravitational field.

As emphasised by \cite[pp.~91--104]{Rovelli2014}, gauge freedom is not a mere artefact of formalism but reflects the relational nature of physical degrees of freedom.\footnote{In this paper, I choose to take Rovelli's position. A different proposal comes from \cite{GomesNoether}, according to whom \lq{}\lq{}gauge symmetry provides a path to building appropriate dynamical theories---and that this rationale invokes the two theorems of Emmy \cite{Noether1918}\rq{}\rq{}. This approach is an extension of the, well-known, answer amongst practising physicists known as the \textit{gauge argument} of \cite{Weyl1929}, which posits that local gauge invariance necessitates the introduction of gauge fields to properly describe fundamental interactions. The topic is broad. I refer the reader to \citep{Teh2015-TEHANO,Weatherall2016-WEAUG}, for some other replies on Rovelli's \lq{}relational proposal on why gauge\rq{}.} Similarly, \cite[p.~3]{Henneaux1994-ka} define a gauge theory as:
\begin{quote}
    [...] a theory in which the dynamical variables are specified with respect to a reference frame.
\end{quote}
Thus, \textbf{DRFs} provide a powerful framework to describe the relational evolution of physical systems by eliminating gauge redundancy and are essential for understanding gauge-invariant observables and diffeomorphism invariance in GR.

In what follows, I characterise a \textbf{DRF} as a material system that allows the deparametrisation of a gauge system purely in terms of dynamically coupled physical fields acting as reference frames.
This also will highlight the role of frame dynamics as a \lq{}gauge-fixing mechanism\rq{} (on this role of reference frames see also \citealp{Brown1995,Thiemann2006, Tambornino2012, BamontiThebault2024}).
In summary, the deparametrisation procedure of a parametrised system enables certain variables to function as reference frames.\footnote{See \citet[Ch.4]{Henneaux1994-ka} for the step-by-step procedure starting from the parametrised action of a Newtonian particle in 1D . Aside from technical differences between the frameworks, the deparametrisation procedure is conceptually analogous to the construction of complete observables as described in \cite{Rovelli_2002} and \cite{Dittrich2006}.}
This insight proves particularly valuable in GR, a theory that \enquote*{naturally} adopts a parametrised form. As Henneaux and Teitelboim note:

\begin{quote}

The already [parametrised] system "per excellence" is the gravitational field in general relativity. \cite[p.102]{Henneaux1994-ka}

\end{quote}

\subsubsection{Four Klein-Gordon Scalar Fields}\label{sec3.2.2}
An informative example of a \textbf{DRF} is provided by a \textit{test} system \citep{Brown1995}. Notably, a test system's back-reaction on gravity is neglected.

As a toy model for a test system, consider a set of four real, massless, free Klein-Gordon scalar fields $\{\phi^{(I)}\}$, coupled with some metric $g_{ab}$ which satisfies the EFEs; for instance, in the absence of other material systems aside from the test one, $R_{ab}=0$ would hold. 

It should be emphasised at the outset that the choice of four Klein–Gordon 
scalar fields is not meant to provide a realistic model of any actual 
material reference system. For that purpose, I will introduce other systems in \S\ref{3.2.2} below. Rather, it serves as an illustrative case that allows one to display in the simplest possible setting how four scalar degrees of freedom can function as a \textbf{DRF},  without introducing unnecessary physical complications. The key point is that the scalar-field framework  is not committed to Klein–Gordon matter as such, but only to the availability of four independent scalar degrees of freedom that can provide the required  spatiotemporal labels.\footnote{Since many fundamental physical fields (whether fermionic or bosonic) are not scalars, these required scalars may represent \enquote*{collective} properties of matter (for example, in FLRW cosmology, the entropy of the cosmological fluid can serve as a reference clock \citep{Schutz1970,Schutz1971,Cianfrani:2009vm,Campolongo2020} or be constructed from fundamental fields.}

Each scalar field satisfies
\begin{equation}
\square_g \phi^{(I)} \equiv \nabla_a \nabla^a \phi^{(I)} = 0,\label{dedonder}
\end{equation}
where \(\nabla_a\) is the covariant derivative compatible with the metric \(g_{ab}\).

In this setup, the dynamics of the scalar fields is coupled to the \textit{given} metric \(g_{ab}\) to define the compatible connection \(\nabla_a\).
Importantly, this case should not be confused to that in which the metric is an \enquote*{absolute} or \enquote*{fixed} background field.
Here, the term \lq{}\textit{given}\rq{} only reflects the assumption that the backreaction of the scalar fields is negligible, consistent with GR's background-independent framework.\footnote{An absolute field is defined as a field which is the same (up to isomorphism) in every DPM.  A fixed field is the same in every KPM \citep{Anderson1967-en}. We must not accept a \textit{given} metric in the above meanings, because that would reduce the discussion to a Klein-Gordon theory in a curved background.  In contrast, here, we deal with GR which is a background independent dynamical theory of the gravitational field (see \cite{James_Read2023-mk,BamontiBI} for an in-depth discussion on this statement of mine, which is not taken for granted). The case examined here is introduced in \citet[\S 8.7]{pooley2022reality} under the name \textbf{GR2}. More generally, one can also have a dynamical metric that takes into account the stress-energy tensor of other material fields, but not that of the reference frame.}

Equation \eqref{dedonder} is analogous to the\textit{ De-Donder gauge-fixing} condition applied to coordinates, demonstrating the correspondence between employing a \textbf{DRF} and imposing gauge conditions in a relational framework. Hence, we have a straightforward example of a gauge-fixing condition, understood \lq{}relationally\rq{} as a set of dynamical equations.\footnote{As stated in \cite{GomesSamediff1}: \lq{}\lq{}Though De Donder gauge is still not complete—it requires initial conditions on the metric and its time derivative (cf. \citealp[p.161]{landsman2021})—it suffices to render evolution deterministic\rq{}\rq{}.}

When the Klein-Gordon fields \(\phi^{(I)}\) are used as the reference frame, one can construct a complete set of local gauge-invariant observables, such as $g_{IJ}(\phi):= \big[ (\phi^{(I)})^{-1} \big]^* g_{ab}$.
In contrast to \textbf{IRFs}, the dynamical coupling between the scalar fields and the metric ensures that once the solution \((g_{ab}, \phi^{(I)})\) is determined, it is \textit{unique}; configurations such as \(([d^*g]_{ab}, \phi^{(I)})\) or \((g_{ab}, d^*\phi^{(I)})\) for generic \(d \in \text{Diff}(M)\) do not represent distinct possible physical solutions. This uniqueness directly reflects the absence of gauge redundancy owing to the dynamical coupling of the \textbf{DRF} with the metric.

\subsubsection{DRFs in the Orthodox View}\label{sec3.2.3}

In much of the literature (see \S \ref{sec2}), a reference frame is often represented by a timelike 4-velocity field \(U^a\), tangent to the congruence of worldlines of test particles or a test matter fluid. 
This orthodox viewpoint, introduced by Earman and Norton \citep{Earman1974, Norton1985} and widely adopted in the physics community \citep{Wald1984, malament2012topics}, treats a reference frame as the expression of matter's state of motion, with a coordinate system \((x^0, x^1, x^2, x^3)\) locally adapted to it.

Under my definition, a \textbf{DRF} is a material system that satisfies its equations of motion and is dynamically coupled to the gravitational field, yet its backreaction on the metric is neglected.\textit{ This naturally accommodates the orthodox view as a special case}.
A comparison between the Earman-Norton conception of reference frames and the notion of \textbf{DRF} provides valuable insights into the distinctions and overlaps between these frameworks. While I do not pursue an exhaustive analysis here, some key differences warrant brief discussion. Recall from \S\ref{sec2}, however, that the four Klein–Gordon scalars of §\ref{sec3.2.1} in general do not, collectively, constitute a 4-velocity.\footnote{Only through the Jacobian $\partial_\mu \phi^{(I)}$ and its invertibility  condition (cf. fn.~\ref{jacobian}) can one construct covectors that 
approximate the role of a tetrad or that define a temporal direction via 
$\nabla_a \phi^{(0)}$. Even then, the resulting objects are derived 
geometrical structures rather than the original scalars themselves. } 
Accordingly, in general the scalar-field \textbf{DRF} should be understood as providing  four independent relational degrees of freedom, not as specifying a  4-velocity in the orthodox sense.
The comparison between the two views is therefore only structural: both approaches anchor physical magnitudes to material systems, but while the orthodox view singles out a timelike direction of motion,  the scalar-field framework assigns four independent coordinates directly.

Notably, while the orthodox approach represents a reference frame by a 4-velocity \(U^a = dx^a/d\tau\) --- with \(\tau\) the proper time of a comoving, but not synchronous observer (see \cite{BamontiThebault2024} for details) --- my framework emphasises a description in terms of four scalar fields that encode spatiotemporal localisation \textit{directly}.

In the orthodox view, the 4-velocity defines the time component along the \textit{direction} of motion, while the spatial components are defined orthogonally.\footnote{The same observation applies to the tetrad frame, where the four tetrads define four spatiotemporal \textit{directions}. Each tetrad, being a 4-vector, does not itself constitute a standard of time or space in the way that a 4-vector defined by four scalar fields does.}

For example, the components of a generic 4-vector are organised as follows:
\begin{itemize}
    \item The \textit{time component} is defined along the \textit{direction} of \( U^a \), which corresponds to the proper time evolution along the worldlines of the fluid.
    \item The \textit{spatial components} are defined orthogonally to \( U^a \), representing \textit{directions} orthogonal to the fluid's motion.
\end{itemize}
This approach retains a coordinate-based framework, as emphasised by Earman and Norton \citep{Earman1974, Norton1985}, who describe such a system as an \enquote*{adapted coordinate system.}

In contrast, the scalar-field approach assigns the full four degrees of freedom via four independent variables, avoiding potential ambiguities arising from coordinate adaptation.
As Earman and Glymour observe:
\begin{quote}
    Of course, a reference frame can be represented by a maximal class of adapted coordinate systems. [...] But such a coordinate representation can easily lead to a blurring of the crucial distinctions [between reference frames and coordinate systems] mentioned above. \cite[p.~254]{Earman1978}
\end{quote}
%This observation underscores the importance of distinguishing between the material system constituting the reference frame and the coordinates used to represent it—a distinction that the \textbf{DRF} framework makes explicit.

To further illustrate, consider a non-backreacting formulation of the \enquote*{Brown-Kuchař dust},  where a dust fluid serves as a reference frame and is represented by eight scalar fields. Among these, four fields—denoted \(T\) and \(Z^i\)—encode the spatiotemporal degrees of freedom \citep{Brown1995}. In this formulation:
\begin{itemize}
    \item The \( T \) field parametrises the proper time along the geodesics.
    \item The \( Z^i \) fields remain constant along the geodesics of the dust.
\end{itemize}
This arrangement enables the dust fluid to function as a global reference frame that foliates spacetime into hypersurfaces of constant \(T\), with the \(Z^i\) fields providing spatial labels within each hypersurface. Crucially, the spatiotemporal localisation is achieved \textit{directly} via these four scalar degrees of freedom, without relying on the dust 4-velocity \(U^a\).

Thus, while the orthodox view ties the notion of a reference frame to the motion of matter as expressed by \(U^a\), my scalar-field approach encodes spatiotemporal localisation directly in the four scalar fields.

In sum, while the orthodox view provides an essential historical and philosophical foundation for understanding reference frames in GR, comparing it with the \textbf{DRF} framework presented here yields a clearer conceptual take of how reference frames are employed and interpreted. This discussion highlights both the strengths and the limitations of the orthodox approach when compared to the broader classification of reference frames proposed in this work.

\subsubsection{Global Positioning System (GPS) Frame}\label{sec3.2.4}

\begin{figure}[h!]
    \centering
    \includegraphics[scale=0.2]{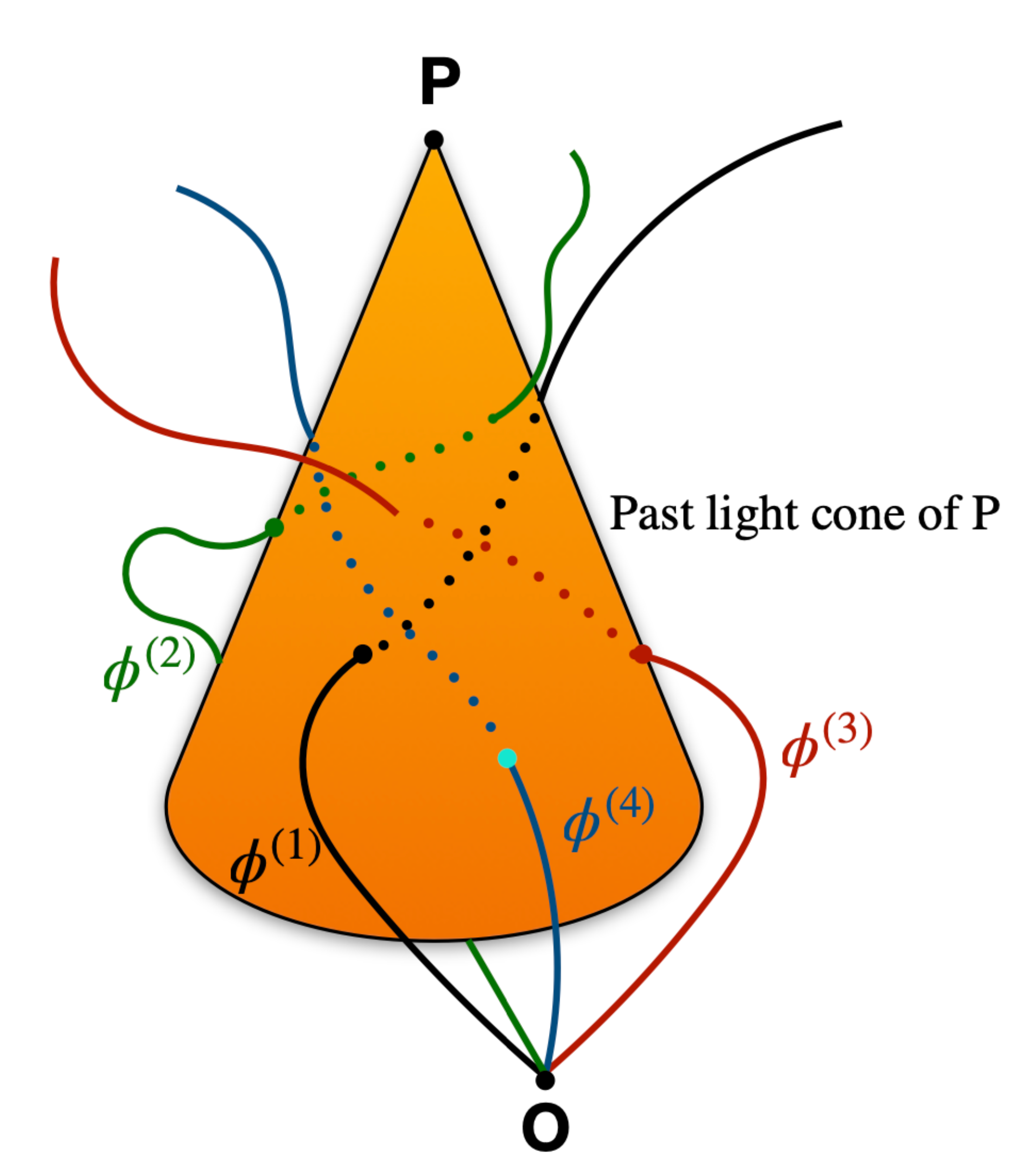}
    \caption{Construction of the set of GPS reference frame. Each scalar $\phi^{(I)}$ corresponds to the proper time of a satellite broadcast at point $P$.}
    \label{fig0}
\end{figure}

To conclude the discussion on \textbf{DRFs}, I argue that a realistic example is provided by the so-called \textit{GPS coordinates}, as introduced in \cite{RovelliGPS}. In this framework, GR is coupled to four test bodies—referred to as satellites—that follow timelike geodesics of a given metric \(g_{ab}\) and are assumed to emanate from a common initial point \(O\).\footnote{Regarding the meaning to be attributed to \lq{}\textit{given}\rq{}, see \S \ref{sec3.2.2} above.}

Each satellite broadcasts its proper time, \(\phi^{(I)}\), and any spacetime point \(P\) can be uniquely associated with four numbers \(\phi^{(I)}(P)\). These numbers, corresponding to the proper timelike distances from \( O \) to the intersections of the past lightcone of \( P \) with the worldlines of the satellites, form the physical variables that define the GPS \textbf{DRF}.  In this way, the GPS system provides a relational framework for spacetime localisation, where the physical metric is expressed not on the manifold \(\mathcal{M}\) but on the space of four-tuples \(\mathbb{R}^4\). See Figure \ref{fig0} for an illustration.

This example not only illustrates the practical implementation of a \textbf{DRF} but also underscores how such a framework addresses key conceptual challenges in GR, including the construction of gauge-invariant local observables and the relational nature of spacetime, as explained in \cite{RovelliGPS}.\footnote{As a further example, also Fletcher’s \enquote*{\textit{light clocks}} ---  which measure proper time \lq{}\lq{}experienced by a point particle along a timelike curve with the length of that curve as determined by the metric\rq{}\rq{} --- can be understood as composing a \textbf{DRF} \cite[p.1369]{Fletcher2013}.}

Having seen how \textbf{DRFs} incorporate dynamics but still neglect stress–energy, we now turn to the fully coupled case—Real Reference Frames.

\subsection{Real Reference Frames (RRFs)}\label{sec3.3}

When both the dynamics of the material reference frame and its stress-energy contribution are fully taken into account, one obtains what I term a Real Reference Frame (\textbf{RRF}). Examples of \textbf{RRFs} include pressureless dust fields considered with their full backreaction \citep{Brown1995} and massless scalar fields \citep{Rovelli1994}. Among the classifications proposed in this work, \textbf{RRFs} are the least mathematically convenient yet the most realistic for modeling reference frames.

%Although \textbf{RRFs} will not be the central focus of the subsequent discussion, it is useful to note that a sub-classification can be introduced. For instance, one may define an \textbf{RRF\textsubscript{dep}} as a Real Reference Frame that allows the theory to be deparametrised, thereby enabling the analytical construction of complete observables \citep{Tambornino2012}. In fact, only in certain cases approximations applied to the Hamiltonian of the material field constituting the \textbf{RRF} can facilitate such a deparametrisation procedure.\footnote{This highlight that even for \textbf{RRFs}, approximations may be necessary, as discussed by \cite{Norton2012}. Similarly, the possibility of globally deparametrising a \textbf{DRF} depends on whether the field used as the reference frame covers the manifold completely—a condition rarely met in GR.}

The detailed study of \textbf{RRFs} is deferred to future work, but their significance is evident in quantum gravity phenomenology. For example, when a quantum material reference frame is in superposition, gravitational backreaction can lead to the splitting of spacetime into distinct branches. As noted by \cite{adlam:2022}:
\begin{quote}
    For the small masses we deal with in current quantum experiments, the difference between the spacetimes is experimentally insignificant, and thus it is typically assumed that we can completely discount any effects of gravitational back-reaction.
\end{quote}
This observation underscores the practical relevance of \textbf{DRFs} in addressing conceptual challenges in GR, while also highlighting the potential importance of \textbf{RRFs} in future developments in quantum gravity.\footnote{For example, \textbf{RRFs} could be useful to explore whether including backreaction effects of the reference frame could lead to observable deviations from the standard Bose-Marletto-Vedral predictions \citep{Bose2017, Marletto2017, QuantumHole}.}

\subsection{Summing Up}

The presented classification provides a clear conceptual framework for addressing key issues in GR that can be solved by treating reference frames as material systems coupled to gravity.
These issues, identified in the Introduction (\S \ref{Introduction}),  are particularly evident when working with uninstantiated coordinate systems or with uncoupled reference frames, such as \textbf{IRFs}. By contrast, using \textbf{DRFs} or \textbf{RRFs} allows for a natural resolution of these problems.

\subsubsection*{Resolution of \textbf{(P1)}: Relational Localisation and Local Gauge-Invariant Observables}

All physically instantiated frames—Idealised (\textbf{IRFs}), Dynamical (\textbf{DRFs}), and Real (\textbf{RRFs})—achieve \textit{relational localisation}, since they tie spacetime points to the values of material degrees of freedom rather than to abstract manifold labels.
However, \textit{only} when the frame’s own dynamics are taken into account (as in \textbf{DRFs} and \textbf{RRFs}) do these relationally localised quantities become genuine complete observables (gauge-invariant).\footnote{On the relationship between relationalism and gauge-invariance see \cite{BamontiGomes2024}. In short, the latter implies the former, but not vice versa.} In an \textbf{IRF}, by contrast, neglecting the dynamical equations of the frame leaves relationally defined objects underdetermined up to arbitrary diffeomorphisms and therefore not gauge-invariant.

\subsubsection*{Resolution of \textbf{(P2)}: Interpretation of Gauge Freedom in GR}

When \textbf{DRFs} or \textbf{RRFs} are used, the gauge-fixing conditions of GR can be interpreted as the dynamical equations of the physical system chosen as the reference frame. In this view, what might be seen as gauge redundancy is reinterpreted as a consequence of neglecting the full dynamics of the reference system (which, if ignored, effectively demotes it to an \textbf{IRF}). As \cite{Rovelli2014} explains:
\begin{quote}
    Gauge invariance is not just mathematical redundancy; it is an indication of the relational character of fundamental observables in physics. [...] Gauge is ubiquitous. It is not unphysical redundancy of our mathematics. It reveals the relational structure of our world. [...] The choice of a particular gauge can be realised \textit{physically} via coupling with a material reference system in general relativity.
\end{quote}
This perspective underscores the relational nature of GR and highlights the central role of material reference systems in the interpretation of gauge invariance.

In conclusion, the proposed classification not only provides semantic clarity regarding the use of reference frames in GR but also enhances our understanding of the relationship between coordinates and reference frames, as I will further discuss in the next section.

\section{The Approximate Modelling of Reference Frames as Coordinates}\label{sec4}

\begin{quote}
    [I]t is not often that experiments are done under the stars. Rather they are done in a room. Although it is physically reasonable that the walls have no effect, it is true that the original problem is set up as an idealization.

    \rightline{Richard Feynman.\footnote{\cite{Feynman1965-qw}. This choice of section opening is also found in \cite{Wallace2022b}.}}
\end{quote}
A notable empirical success of GR is the detection of gravitational waves by the LIGO project \citep{LIGO}. 
In this context, the gravitational contribution of the reference frame used to localise the detection of gravitational waves on Earth is entirely disregarded. 
The metric components are calculated within a specific coordinate gauge—the Transverse-Traceless (TT) gauge—illustrating how, in the process of modelling measurement outcomes, reference frames are often modelled in ways that render them equivalent to coordinate systems.\footnote{Local coordinate systems are routinely employed to solve the Einstein Field Equations. For instance, Schwarzschild coordinates \((t, r, \theta, \phi)\) provide a convenient description of the Schwarzschild geometry. Of course, this geometry can be expressed in various coordinate systems, highlighting the arbitrariness of such choices.
In traditional textbook interpretations, Schwarzschild coordinates are often understood as labels for spatiotemporal points, without any connection to a physical system instantiating them. In the sense discussed by \cite{Norton2012}, such variables are idealisations.}

This functional should not be read as a limitation of experimental or theoretical practice, but rather as the natural outcome of approximation procedures involved in modelling physical systems. 
Being aware of such procedures and that material reference frames are implicitly employed, however, offers critical insight into the presence of gauge freedom and the physical role of gauge-fixing conditions. 
As discussed in \S \ref{sec2}, diffeomorphism invariance—a form of gauge redundancy—implicitly relies on an approximation procedure that neglects the dynamics of the physical system serving as the reference frame. 
For example, in the analysis of gravitational waves, the TT-gauge conditions can be understood relationally as a set of dynamical equations satisfied by the reference frame that localises measurements in space and time. 
This correspondence between gauge-fixing and the role of reference frames provides a constructive way to articulate the conceptual foundations of the theory.

A plausible narrative is as follows.

In the process of modelling measurement outcomes, the reference frame is approximated to such an extent that it effectively becomes what I have defined as an Idealised Reference Frame (\textbf{IRF}). 
This approximation procedure leads to treating the reference frame \textit{as if} it were nothing more than a coordinate system.
Essentially, both \textbf{IRFs} and coordinate systems involve non-dynamical variables that serve to define the spatiotemporal localisation of relevant quantities. And both do not allow to define local gauge-invariant observables, unless some gauge-fixing is explicited.
Yet the two remain distinct in principle. 
In particular:
\begin{itemize}
    \item \textbf{Coordinates} are mathematical, uninstantiated labels that, \textit{by definition}, lack dynamics. They are \emph{idealisations} in the sense of \cite{Norton2012} and do not correspond to any physical system.
    \item \textbf{IRFs} are \textit{approximations} of real, material systems within the theory's model. These systems, while physically instantiated and interacting with other degrees of freedom, are subject to dynamical approximations that \lq{}demote\rq{} their role—effectively making them behave like coordinate systems. Moreover, unlike coordinates, \textbf{IRFs} are conceived as fields that covary under active diffeomorphisms (see \citealp{BamontiGomes2024}). The approximation procedure cannot neglect the reference frame's field nature.
\end{itemize}

This distinction between idealisations (coordinates) and approximations (\textbf{IRFs}) is at the heart of their different role in the theory. 
As Norton argues, the difference between idealisations and approximations \emph{matters} because it distinguishes entities that are physically instantiated from those that are purely mathematical artefacts.

Ultimately, distinguishing reference frames from coordinate systems clarifies which elements of a theory represent physical systems and which are merely mathematical constructs. As \cite{Maudlin2018} notes, philosophers and theoretical physicists often fail to differentiate between these two aspects. Similarly, \cite{GomesHowChooseGauge} emphasise:
\begin{quote}
    No special care is taken to specify: which parts represent ontology, ‘what there is’ $[\dots]$ and which parts represent nothing physical, but instead mathematics (which, though unphysical, can of course be invaluable for calculation).
\end{quote}
My classification and field-theoretic approach contribute to this clarification and may help shape a coherent interpretation of physical theories.\footnote{See also \cite{Curielschema}, where Curiel argues that to fully understand the structure and nature of knowledge in physics, it is necessary to analyse how observers, measuring instruments and experimental set-ups are modelled within the theory itself. It is therefore necessary to \lq{}schematise the observer\rq{}.}

\section{Conclusion}\label{sec6}

This work introduced a novel three-fold classification of material reference frames in GR, distinguishing between Idealised Reference Frames (\textbf{IRFs}), Dynamical Reference Frames (\textbf{DRFs}), and Real Reference Frames (\textbf{RRFs}).
 
This hierarchy formed a robust framework for tackling two foundational challenges in GR: the challenge of defining local \textit{and} gauge-invariant observables \textbf{(P1)}, and the challenge of providing a physical interpretation to diffeomorphism gauge freedom \textbf{(P2)}. Specifically, the use of \textbf{DRFs} and \textbf{RRFs} resolve both challenges.
This contrasts with \textbf{IRFs}, which enable relational localisation but they behave like coordinate systems, as far as dynamics is concerned. While \textbf{IRFs} are often sufficient for modelling purposes, they may underrepresent the relational character of GR and physical interactions.

By contextualising reference frames in GR, this work complements existing literature and clarifies the distinction between coordinates and reference frames. Coordinates are mathematical idealisations—uninstantiated variables that exist purely within the formalism—while reference frames are physically instantiated structures, often approximated by \textbf{IRFs}.
In this sense, the proposed three-fold classification is explanatory rather than purely terminological: it shows how different practices within the standard literature—such as the use of congruences of test particles, the adoption of scalar fields, or the GPS implementation of reference frames—can be located within a single unified framework.

Beyond these conceptual gains, the proposed classification has potential implications for the emerging study of quantum reference frames. When material systems are treated as quantum reference frames—capable of being in superposition and subject to gravitational backreaction—the resulting spacetime dynamics demands new frameworks for understanding gravitational interactions in the quantum regime. New investigations, such as those in \cite{QuantumHole}, have already begun to explore these issues within the context of the present framework.

In summary, the systematic classification of reference frames presented in this work not only addresses key challenges offering a novel framework, but also enriches our understanding of the interplay between matter and geometry in Einsteinian gravity, paving the way for further exploration in both classical and quantum gravitational contexts.

\clearpage

\bibliography{BIB2.bib}

\end{document}